\documentclass[12pt]{iopart}
\usepackage{graphicx,iopams}

\begin{document}

\title{Dynamical properties of dipolar Fermi gases}

\date{\today}
\author{T Sogo$^1$, L He$^2$, T Miyakawa$^3$, S Yi$^2$, H Lu$^4$ and H Pu$^4$}

\address{$^1$Institut f\"ur Physik, Universit\"at Rostock,
D-18051 Rostock, Germany}
\address{$^2$Institute of Theoretical Physics, Chinese Academy of Sciences, Beijing
100190, China}
\address{$^3$Department of Physics, Faculty of Science,
Tokyo University of Science, 1-3 Kagurazaka, Shinjuku, Tokyo, 162-8601, Japan}
\address{$^4$Department of Physics and Astronomy,
and Rice Quantum Institute, Rice University, Houston, Texas 77251-1892, USA}

\begin{abstract}
We investigate dynamical properties
of a one-component Fermi gas with dipole-dipole interaction between particles.
Using a variational function
based on the Thomas-Fermi density distribution
in phase space representation,
the total energy is described by
a function of deformation parameters in both
real and momentum space.
Various thermodynamic quantities of a uniform dipolar Fermi gas
are derived, and then instability of this system is discussed.
For a trapped dipolar Fermi gas, the collective
oscillation frequencies are derived with the energy-weighted sum rule method.
The frequencies for the monopole and quadrupole modes are calculated,
and softening against collapse is shown as the dipolar strength approaches the critical value.
Finally, we investigate the effects of the dipolar interaction on the expansion dynamics of the Fermi gas and show how the dipolar effects manifest in an expanded cloud.
\end{abstract}


\submitto{\NJP}


\maketitle

\section{Introduction}

In recent years, atomic quantum dipolar gases have received much interest, for the simple reason that the anisotropic and long-range nature of the dipole-dipole interaction gives rise to a rich spectrum of novel properties to
such systems. The theoretical study of dipolar Bose-Einstein condensates started in 2000. Properties of ground state~\cite{ssz00,you}, collective oscillations~\cite{you1,gs02}, topological defects such as spin textures and vortex states~\cite{ksu06,yp06} are studied. Moreover, when confined in optical lattice potentials,
various quantum phases, such as
ferromagnetism~\cite{pzm01}, and
supersolid state~\cite{gsl02,syis}, etc. are predicted.
Theoretical studies of dipolar Fermi gas
have been carried out for ground state~\cite{ger01},
excitatations~\cite{gbr03},
BCS superfluidity~\cite{bmr02} and rotating properties~\cite{rotate}. A recent review of dipolar quantum gases can be found in Ref.~\cite{review}.

In experiments,
Bose-Einstein condensation of chromium atoms,
which possess a magnetic dipole moment six times larger than that of alkali atoms,
have been realized~\cite{gwh05,Cr}.
The effect of dipole-dipole interaction
in $^{52}$Cr condensate is observed
in its expansion dynamics~\cite{lkf07}.
Besides chromium, heteronuclear molecules~\cite{mtc04,szs04,igo04,wqs04,ooh06,junye}
and Rydberg atoms~\cite{g94,cgp05,dkh08}
are also expected to interact via strong dipole-dipole force due to their large electric dipole moment,
and their experimental realization is under way in a number of groups.

In Ref.~\cite{msp08},
three of us studied the ground state properties
of a dipolar Fermi gas
by employing a variational Wigner function based on the
Thomas-Fermi density of identical fermions.
We showed that the dipole-dipole interaction
induces a deformation of the
momentum space distribution,
and identified that such deformation arises from the Fock exchange term, which had not been paid particular attention in previous studies. The purpose of this
paper is to extend the work of ref.~\cite{msp08},
and investigate the collective excitations and
expansion dynamics of the dipolar Fermi gas. We want to emphasize that, due to the Pauli exclusion principle, the energy scales of a fermionic system is much larger than those of a Bose condensate. Consequently, the dipolar effects in Fermi gas only becomes significant when the dipole moment is very large. Our calculations show that for heteronuclear molecules with typical electric dipole moment on the order of one Debye, dipolar effects can be easily detected. While dipolar effects are usualy negligible in atomic Fermi gases \footnote{As pointed out in Ref.~\cite{review}, the magnetic dipole moment of chromium is equivalent to an electric dipole moment of 0.056 Debye.}.

The content of the paper is organized as follows.
In the next section,
we present the model Hamiltonian and
the total energy of the one-component dipolar Fermi gas
under Hartree-Fock approximation.
In section \ref{sec-homo},
we derive the total energy function
in a uniform system with a variational ansatz
of Fermi surface and compute various thermodynamic quantities
of the system. Here we show how the Fock exchange interaction leads to Fermi surface deformation as well as the instability of the system.
In section \ref{sec-trap}, we turn our attention to a trapped system and investigate
various modes of collective excitations using
the sum-rule method, and show the softening of
the excitation frequency as the interaction strength is increases towards a critical value.
In section \ref{sec-expansion}, we study the expansion dynamics of an initially trapped Fermi gas and show how the expanded cloud bears the signature of the underlying dipolar interaction.
Finally, a
summary is presented in section \ref{sec-summary}.

\section{\label{sec-tef} Total energy functional in phase space representation}
We consider
a single component Fermi gas of atoms or molecules with
dipole moment aligned along the axial axis of a cylindrical harmonic trap.
The Hamiltonian of this system is described by
\begin{eqnarray}
\hat H=\sum_i\left[
-\frac{\hbar^2}{2m}\vec \nabla_i^2
+\frac{1}{2}m\left\{\omega_\rho^2 (x_i^2+y_i^2)+\omega_z^2z_i^2\right\}\right]
+\sum_{i>j}V_{dd}({\bf r}_i -{\bf r}_j),
\label{H}
\end{eqnarray}
where $m$ is the mass of fermions, and
$\omega_\rho$ and $\omega_z$ are the oscillation frequencies
along the radial and axial axes, respectively.
The dipole-dipole interaction of the last term in Eq.~(\ref{H})
is described by
$V_{dd}({\bf  r})=d^2(1-3\cos^2\theta)/r^3$,
where $\theta$ is the angle between ${\bf r}$ and
the dipole moment ${\bf d}$.

In the Hartree-Fock approximation,
the total energy derived from Hamiltonian (\ref{H})
can be written as the sum of the kinetic, trapping potential, Hartree direct and Fock exchange energies 
\begin{eqnarray}
E&=&E_{kin}+E_{ho}+E_{d}+E_{ex} \label{et}\\
E_{kin}&=&\int \!\! d^3 r  \!\! \int \!\! \frac{d^3 k}{(2\pi)^3}\,
\frac{\hbar^2k^2}{2m}f({\bf r},{\bf k}) \label{ekin-ps}\\
E_{ho}&=& \int \!\! d^3 r \!\! \int \!\! \frac{d^3 k}{(2\pi)^3}\,
\frac{1}{2}m[\omega_\rho^2(x^2+y^2)+\omega_z^2z^2]\,
f({\bf r},{\bf k}) \label{eho-ps}\\
E_{d}&=&\frac{1}{2}\int \!\! d^3 r \!\! \int \!\! d^3 r' \!\!
\int \!\! \frac{d^3 k}{(2\pi)^3}\int \!\! \frac{d^3 k'}{(2\pi)^3}\,
V_{dd}({\bf r} -{\bf r}')f({\bf r},{\bf k}) f({\bf r}',{\bf k}') \label{ed-ps}\\
E_{ex}&=&-\frac{1}{2}\int \!\! d^3 R \!\! \int \!\! d^3 s
\!\! \int \!\! \frac{d^3 k}{(2\pi)^3}\!\! \int \!\! \frac{d^3 k'}{(2\pi)^3}\,
V_{dd}({\bf s})
e^{i({\bf k}-{\bf k}')\cdot {\bf s}}f({\bf R},{\bf k}) f({\bf R},{\bf k}')\,,
\label{eex-ps}
\end{eqnarray}
where
we have introduced the Wigner function
$f({\bf r},{\bf k})$
defined by the following transformation:
\begin{eqnarray}
n({\bf r},{\bf r}')
=\frac{1}{(2\pi)^3}\int \! d^3k \,e^{i{\bf k} \cdot ({\bf r} - {\bf r}')}
f\left(\frac{{\bf r} + {\bf r}'}{2},{\bf k}\right),
\end{eqnarray}
where the one-body density matrix $n({\bf r},{\bf r}')=
\sum_{\alpha}\psi_\alpha({\bf r})\psi_\alpha^*({\bf r}')$ is defined in terms of a complete set of
single-particle wave function $\{ \psi_{\alpha}({\bf r}) \}$.
In Eq.~(\ref{eex-ps}), we have introduced the center of mass coordinate ${\bf R}=({\bf r} + {\bf r^\prime})/2$
and relative coordinate ${\bf s} = {\bf r}-{\bf r}^\prime$.
For the ground state, the summation over single-particle states $\alpha$ goes
from the lowest one up to the Fermi energy.

In our work, we do not calculate the Hartree-Fock energy represented by
Eq.~(\ref{et}) in a fully self-consistent manner, which will be a quite complicated task.
Instead, we adopt a much simpler semiclassical approach and calculate the total energy
by employing a variational ansatz for the Wigner distribution function based on the Thomas-Fermi approximation,
which assumes that the local Fermi surface has the same form
as in homogeneous case at each spatial point. The ground state is then obtained by optimizing the Wigner function that minimizes the total energy. The details of this calculation can be found in Ref.~\cite{msp08}. In the present paper, we will focus on the dynamical properties such as the low-lying collective excitations and the expansion dynamics of the ground state.

\section{\label{sec-homo}Equilibrium properties
of a homogeneous dipolar Fermi gas}

It is instructive to first consider a homogeneous system
($\omega_\rho=\omega_z=0$) in a large box of volume
${\cal V}(=\int d^3 r)$ with number density $n_f$, as this will provide important insights into the trapped system to be studied later.

We introduce the following number-conserving variational ansatz for the Wigner function
\begin{eqnarray}
f({\bf k})
=\Theta\Bigl(k_F^2-\frac{k_\rho^2}{\alpha}-\alpha^2k_z^2
\Bigr),\label{msdtf}
\end{eqnarray}
where $\Theta(\cdot)$
is Heaviside's step function, $k_\rho^2=k_x^2+k_y^2$, and $k_F=(6\pi^2n_f)^{1/3}$
corresponds to the Fermi momentum.
The parameter $\alpha$ characterizes the deformation of the Fermi surface:
$\alpha > 1$ ($<1$) corresponds to an oblate (prolate) Fermi surface. The physical origin of the Fermi surface deformation can be attributed to the anisotropic nature of the dipolar interaction.

\begin{figure}
\begin{center}
\includegraphics[width=70mm]{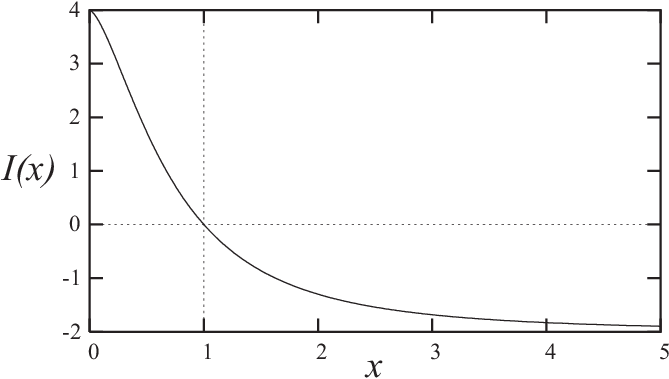}
\end{center}
\caption{
\label{fig-deform}
Deformation function $I(x)$ as a function of $x$.}
\end{figure}

Given the ansatz Eq.~(\ref{msdtf}), the total energy of the homogeneous system can be derived as
\begin{equation}
\varepsilon(\alpha)=E/{\cal V}
=\frac{\hbar^2}{m}
n_f^{5/3}\left[C_1\left(\frac{2\alpha}{3}+\frac{1}{3\alpha^2}\right)
-\frac{\pi}{3}C_{dd}I(\alpha)\right]
\label{ethomo}
\end{equation}
where $C_1=3(6\pi^2)^{2/3}/10$, $C_{dd}=md^2 n_f^{1/3}/\hbar^2$ is the dimensionless dipolar interaction strength,
and
\begin{eqnarray*}
I(x)
&=&\int^\pi_0 d\theta \sin\theta
\Bigl(\frac{3\cos^2\theta}{x^3\sin^2\theta+\cos^2\theta}-1\Bigr)
\nonumber \\
&=&
\left\{
\begin{array}{ll}
\displaystyle
\frac{6}{1-x^3}
\Bigl[1-\sqrt{\frac{x^3}{1-x^3}}
\arctan\bigl(\sqrt{\frac{1-x^3}{x^3}}\bigr)\Bigr]-2 & (x < 1)
\vspace{4mm}\\

\displaystyle
0 & (x=1) \vspace{2mm}\\

\displaystyle
\frac{6}{1-x^3}
\Bigl[1+\frac{1}{2}\sqrt{\frac{x^3}{x^3-1}}
\log\bigl(\frac{\sqrt{x^3}+\sqrt{x^3-1}}{\sqrt{x^3}-\sqrt{x^3-1}}
\bigr)\Bigr]-2 & (x > 1)
\end{array}
\right.
\end{eqnarray*}
is the ``deformation function'' \cite{msp08} and is illustrated in Figure \ref{fig-deform}. $I(x)$ decreases
monotonically from 4 to $-2$ as $x$ increases from 0 to $\infty$, and passes through zero at $x=1$. The first and the second term in the square bracket of Eq.~(\ref{ethomo}) represent the kinetic and the Fock exchange energy, respectively. For the homogeneous system, the Hartree direct energy vanishes.

\begin{figure}
\begin{center}
\includegraphics[width=70mm]{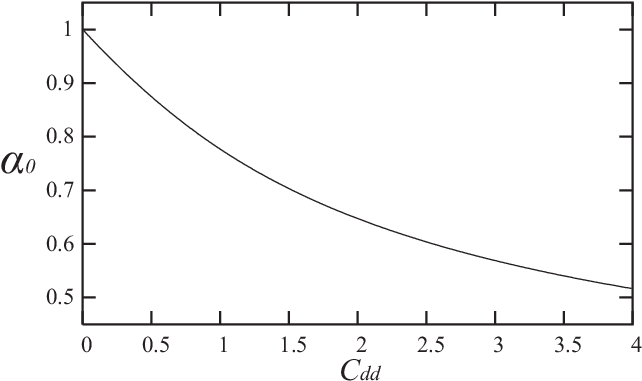}
\end{center}
\caption{\label{fig-a0vscdd}
Optimal deformation parameter $\alpha_0$
as a function of $C_{dd}$. For a molecular Fermi gas with electrical dipole moment $d=1$ Debye, molecular mass $m=100$ a.m.u. and density $n_f = 10^{13}$ cm$^{-3}$, we have $C_{dd} \approx 3.2$.}
\end{figure}

Under this variational approach, the ground state is determined by the stationary condition
for the total energy of Eq.~(\ref{ethomo}) with respect to
parameter $\alpha$:
$\left[d\varepsilon/d\alpha \right]_{\alpha=\alpha_0}=0$. The optimal value $\alpha_0$ is shown in Figure \ref{fig-a0vscdd} as a function of the dipolar strength $C_{dd}$. For free fermion systems, momentum density distribution
is spherical, i.e., $\alpha_0=1$ at $C_{dd}=0$.
As the interaction strength increases,
$\alpha_0$ decreases, which means that the momentum density distribution
becomes more prolate in shape. In other words, the Fermi surface is stretched along the direction of the dipoles.

Once we have the energy of the system as represented by Eq.~(\ref{ethomo}), we can easily obtain other important thermodynamic quantities. Here we provide our calculation for the pressure $P$, compressibility $K$ and chemical potential $\mu$:
\begin{eqnarray*}
P &=& -\left. \frac{\partial E}{\partial {\cal V}}\right|_N = \frac{n_f}{\cal V} \left.  \frac{\partial E}{\partial n_f}\right|_N =\frac{C_{dd}}{3 {\cal V}}  \left.  \frac{\partial E}{\partial C_{dd}}\right|_N \\
\frac{1}{K} &=& n_f \, \frac{\partial P}{\partial n_f } = \frac{C_{dd}}{3}\, \,\frac{\partial P}{\partial C_{dd} } \\
 \mu &=& \left. \frac{\partial E}{\partial N} \right|_{\cal V} = \frac{1}{\cal V}\left. \frac{\partial E}{\partial n_f} \right|_{\cal V}=\frac{1}{N}\,(E+P{\cal V})
\end{eqnarray*}

\begin{figure}
\begin{center}
\includegraphics[width=70mm]{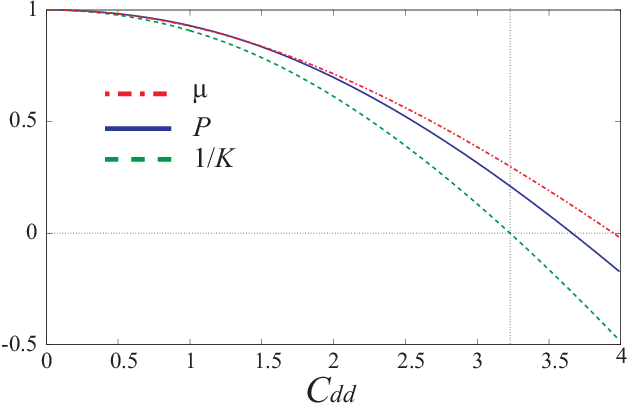}
\end{center}
\caption{\label{fig-comp}
Chemical potential $\mu$, pressure $P$, and inverse compressibility or bulk modulus $1/K$ 
as functions of $C_{dd}$. All quantities are normalized to their corresponding values in the non-interacting limit: $\mu_0=(5C_1/3) \hbar^2 n_f^{2/3}/m =E_F$, $P_0=(2C_1/3) \hbar^2 n_f^{5/3}/m$, and $1/K_0=(10C_1/9) \hbar^2 n_f^{5/3}/m$. The vertical line indicates the critical dipolar strength beyond which the system becomes unstable against collapse.}
\end{figure}
These quantities are illustrated in Figure~\ref{fig-comp}. One can see that $P$, $1/K$ and $\mu$ all monotonically decrease as the dipolar interaction strength increases. In particular, when the inverse compressibility (i.e., the bulk modulus) becomes negative, the system is no longer stable against collapse. Our calculation indicates that the critical dipolar strength is about $C_{dd}=3.23$.


\section{\label{sec-trap}Collective oscillations
of trapped dipolar Fermi gas}

Let us now turn our attention to the trapped dipolar Fermi gas.

First,
to obtain the the total energy of Eq.~(\ref{et}),
we introduce the following ansatz for the Wigner function:
\begin{eqnarray}
f({\bf r}, {\bf k})
=\Theta\left(k_F^2-\frac{k_\rho^2}{\alpha}-\alpha^2k_z^2
-\frac{\lambda^2}{a_{ho}^4}\left(\beta\rho^2+\frac{z^2}{\beta^2}\right)
\right),\label{psdtf}
\end{eqnarray}
where $\rho^2=x^2+y^2$, and
$a_{ho}=\sqrt{\hbar/(m\omega)}$ with $\omega=(\omega_\rho^2\omega_z)^{1/3}$.
The variables $\beta$ and $\lambda$ represent the deformation and
compression of the spatial density distribution
of the system, respectively.
When we take $\alpha=1$,  $\beta=(\omega_\rho/\omega_z)^{2/3}$,
and $\lambda=1$, this trial function is consistent
with the Thomas-Fermi density
of a free Fermi gas in the harmonic trap.
Fermi wave number $k_F$ is related to the number of fermions as
\begin{eqnarray}
N=\int d^3 r \, n({\bf r})=
\int d^3 r \int \frac{d^3 k}{(2\pi)^3} f({\bf r},{\bf k})
=\frac{a_{ho}^6k_F^6}{48\lambda^3}.
\label{pnum}
\end{eqnarray}
%

Substituting Eq.~(\ref{psdtf}) into
Eqs.~(\ref{ekin-ps}), (\ref{eho-ps}), (\ref{ed-ps}), and (\ref{eex-ps}),
we obtain the total energy in units of $N^{4/3}\hbar\omega$ as \cite{msp08}
\begin{eqnarray}
\epsilon(\alpha,\beta,\gamma)
=\frac{E}{N^{4/3}\hbar\omega}
=\epsilon_{kin}(\alpha,\lambda)
+\epsilon_{ho}(\beta,\lambda)
+\epsilon_{d}(\beta,\lambda)
+\epsilon_{ex}(\alpha,\lambda)
\label{ettrap}\\
\epsilon_{kin}
=c_1\lambda\left(2\alpha+\frac{1}{\alpha^2} \right), \quad \,\,\;\;\;
\epsilon_{ho}
=\frac{c_1}{\lambda}\left(\frac{2\beta_0}
{\beta}+\frac{\beta^2}{\beta_0^2}\right) \label{ehotrap}\\
\epsilon_{d}
=N^{1/6} c_{dd} c_2 I(\beta)\lambda^{3/2},\qquad
\epsilon_{ex}
=-N^{1/6} c_{dd} c_2 I(\alpha)\lambda^{3/2}
\label{eextrap}
\end{eqnarray}
where $\beta_{0}=(\omega_\rho/\omega_z)^{2/3}$ represents the trap aspect ratio,
$c_1=3^{1/3}/2^{8/3}\simeq 0.2271$,
$c_2=2^{10}/(3^{7/2}\cdot 5 \cdot 7 \pi^2)\simeq 0.0634$,
and $c_{dd}=d^2/(\hbar\omega a_{ho}^3)$ is the dimensionless dipolar interaction strength for the trapped system. The momentum space deformation parameter $\alpha$, as in the homogeneous case, appears only in the kinetic and the exchange energy terms, both of which are independent of the spatial deformation parameter $\beta$. This indicates that the momentum space distribution of the trapped system will also be elongated along the direction of the dipoles, regardless the geometry of the trapping potential. On the other hand, $\beta$ appears only in the potential energy and the Hartree direct energy terms.

\begin{figure}
\begin{center}
\includegraphics[width=130mm]{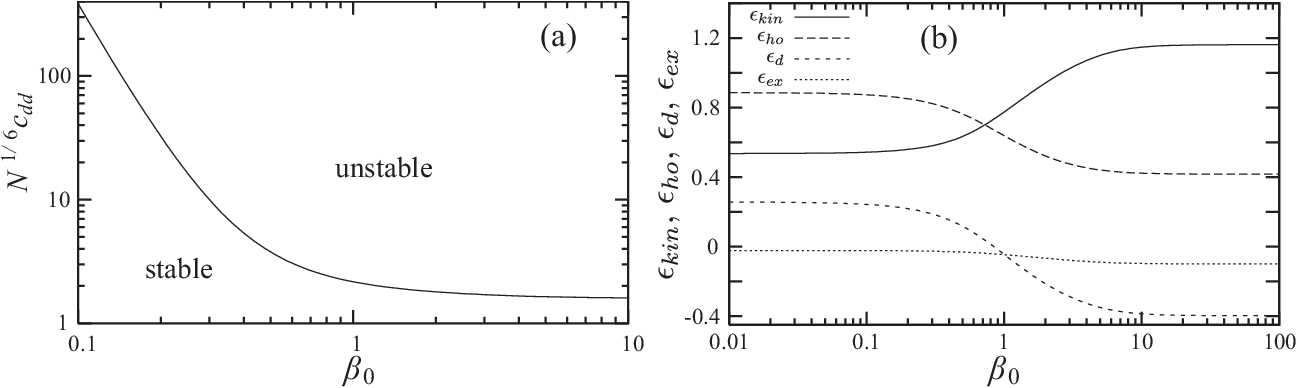}
\end{center}
\caption{\label{stability}
(a) Stability phase diagram in the space of the trap aspect ratio and the dipolar interaction strength. (b) Different energy terms, in units of $N^{4/3}\hbar \omega$ as functions of the trap aspect ratio $\beta_0$
at $N^{1/6}c_{dd}=1.5$.}
\end{figure}

The ground state is determined by
the stationary condition for Eq.~(\ref{ettrap})
with respect to the three variables $\alpha$, $\beta$ and $\lambda$:
$\partial \epsilon/\partial \alpha
=\partial \epsilon/\partial \beta
=\partial \epsilon/\partial \lambda=0$. From the last condition, we can see that the energies of the dipolar Fermi gas satisfy the Virial theorem:
\[2\epsilon_{kin}-2\epsilon_{ho}+3(\epsilon_{d}+\epsilon_{ex})=0 \]
In addition, the ground state has to satisfy
the stability condition:
The energy surface in the coordinates $(\alpha,\beta,\lambda)$
has to be a convex downward
at the stationary point.
If no values of $(\alpha,\beta,\lambda)$ can be found to satisfy both
the stationary and the stability conditions,
the systems is considered to be unstable against collapse~\cite{msp08}. This procedure leads to the stability phase diagram as shown in Figure ~\ref{stability}(a). Just as in the homogeneous case, the trapped dipolar gas is only stable for dipolar interaction strength below a critical value. In Figure ~\ref{stability}(b), we show the different energy terms [Eqs.~(\ref{ehotrap}) and (\ref{eextrap})] as functions of $\beta_0$ at $N^{1/6}c_{dd}=1.5$. Several features are worth pointing out: (1) The exchange energy is always negative, as in the homogeneous case, regardless of the trap geometry; whereas the sign of the direct energy
$\epsilon_{d}$ depends on trap geometry: $\epsilon_{d}>0$
for $\beta_0\lesssim 1$ (oblate trap) and $\epsilon_{d}<0$
for $\beta_0\gtrsim 1$ (prolate trap). (2) Both the kinetic and the trapping energies depend on trap aspect ratio. By contrast, for non-interacting system, when expressed in the same units, we have $\epsilon_{kin} =\epsilon_{ho}= 3c_1 \approx 0.68$ independent of $\beta_0$.

Next, we derive the collective oscillation frequency for several low-lying excitation modes
of the system using the sum rules
in the present formulation~\cite{blm79,ls89}.
In this approach,
we represent the excitation frequency $\Omega$
using the first and third energy-weighted moments of the strength function for a given transition operator $\hat F$:
\begin{eqnarray}
\hbar\Omega = \sqrt{\frac{S_3}{S_1}} \label{sumrule} \\
S_1
=
\sum_{\nu\neq0} (E_\nu-E_0) |\langle \nu | \hat F | 0 \rangle|^2
=\frac{1}{2}\langle 0 |[\hat F,[\hat H,\hat F]]| 0 \rangle \label{firstsum}\\
S_3
=
\sum_{\nu\neq0} (E_\nu-E_0)^3 |\langle \nu | \hat F | 0 \rangle|^2
=-\frac{1}{2}
\langle 0 |[[H,F],[H,[H,F]]]| 0 \rangle,\label{thirdsum}
\end{eqnarray}
where $|\nu\rangle$ denotes the $\nu$-th eigenstate of the Hamiltonian
with eigenenergy $E_\nu$.

For our purpose,
we choose the one-body operator as
\begin{eqnarray}
\hat F=\sum^N_{i=1} F(\vec r_i)=\sum^N_{i=1}
\left[\xi(x_i^2+y_i^2) + \zeta z_i^2 \right], \label{obo}
\end{eqnarray}
where $\xi$ and $\zeta$ are certain parameters.
A collective oscillation is compressive
when $\xi$ and $\zeta$ have the same sign,
and quadrupolar
when they have opposite signs.
The natural monopole and quadrupole operators
correspond to $\xi/\zeta=1$ and $\xi/\zeta=-1/2$, respectively.

Using Eq.~(\ref{obo}), the collective excitation frequency
$\Omega$ in Eq.~(\ref{sumrule}) in the present formulation
can be shown to be
\begin{eqnarray}
\fl
\quad\frac{\Omega}{\omega}
=\sqrt{
\frac{4(\epsilon_{ho\rho}\xi^2+\epsilon_{hoz}\zeta^2)
+{\cal A}(4\xi-\zeta)\zeta
+{\cal B}(4\xi-\zeta)(\xi-\zeta)
+{\cal C}(\xi-\zeta)^2}
{\displaystyle  { \epsilon_{ho\rho}}\xi^2/\beta_0
+\beta_0^2\epsilon_{hoz}\zeta^2}}  \label{Omega}\\
\fl
\quad{\cal A}=\frac{1}{2}(\epsilon_{d}+\epsilon_{ex}), \qquad
{\cal B}=\frac{2}{9}
 \left(\!\beta\frac{\partial \epsilon_{d}}{\partial \beta}
    \!+\!\alpha\frac{\partial \epsilon_{ex}}{\partial \alpha}\right),\qquad
{\cal C}=\frac{2}{9}
 \!\left(\!\beta^2\frac{\partial^2 \epsilon_{d}}{\partial \beta^2}
    \!+\!\alpha^2\frac{\partial^2 \epsilon_{ex}}{\partial \alpha^2}\!\right)
\label{freq}
\end{eqnarray}
where $\epsilon_{ho\rho}=2c_1\beta_0/(\lambda\beta)$ and $\epsilon_{hoz}=c_1\beta^2/(\lambda\beta_0^2)$ are
the radial and axial components of the trapping energy, respectively
[see Eq.~(\ref{ehotrap})].
The excitation frequency $\Omega$ is calculated
by substituting the variational parameters $(\alpha,\beta,\lambda)$
at the stationary point of the total energy (\ref{ettrap}).

From Eq.~(\ref{Omega}) we can easily find the excitation frequencies of
the monopole and quadrupole modes, which have the following expressions:
\begin{eqnarray}
\Omega_M &=& \omega\sqrt{\frac{4\epsilon_{ho}+3{\cal A}}
{{\epsilon_{ho\rho}}/{\beta_0}+\beta_0^2\epsilon_{hoz}}}\,, \label{freq-M}
\\
\Omega_Q &=& \omega\sqrt{\frac{4\epsilon_{ho\rho}+16\epsilon_{hoz}
-12{\cal A}+6{\cal B}+9{\cal C}}
{{\epsilon_{ho\rho}}/{\beta_0}+4\beta_0^2\epsilon_{hoz}}}\,.
\label{freq-mq}
\end{eqnarray}
The corresponding frequencies for non-interacting system can be recovered
from Eqs.~(\ref{freq-M}) and (\ref{freq-mq}) as
$\Omega_{M0}=\sqrt{12\omega_\rho^2\omega_z^2/(\omega_\rho^2+2\omega_z^2)}$ and
$\Omega_{Q0}=\sqrt{12\omega_\rho^2\omega_z^2/(2\omega_\rho^2+\omega_z^2)}$.

\begin{figure}
\begin{center}
\includegraphics[width=130mm]{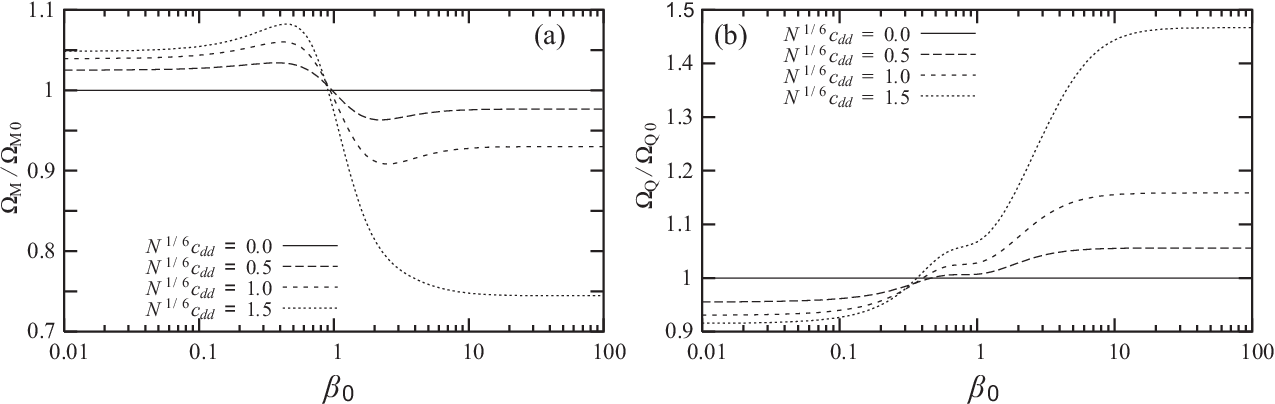}
\end{center}
\caption{
\label{fig-ex-mono}
Excitation frequencies of the monopole mode $\Omega_M$ ($\xi=\zeta=1$) and the quadrupole mode $\Omega_Q$ ($\xi=1$, $\zeta=-2$) as a function of $\beta_0$.
The frequencies are normalized to the corresponding values of the non-interacting system.}
\end{figure}

Figure \ref{fig-ex-mono} shows the excitation frequency
of the monopole mode $\Omega_M$ and the quadrupole mode $\Omega_Q$. As can be seen in Figure~\ref{stability}(b), the total interaction energy $(\epsilon_{ex}+\epsilon_{d})$ is positive (in other words, the overall dipolar interaction is repulsive) for oblate traps ($\beta_0 <1$) which makes the atomic cloud more compressible, hence $\Omega_M$ is increased compared to its non-interacting values. For prolate traps ($\beta_0 >1$), the opposite will be true. This is consistent with the result shown in Figure \ref{fig-ex-mono}(a). The quadrupole mode frequency $\Omega_Q$, on the other hand, exhibits a roughly opposite trend.

To account for the hybridization of different modes, we parameterize $\xi$ and $\zeta$ in Eq.~(\ref{obo}) as
$\xi=\sin\theta$ and $\zeta=\cos\theta$, with $0 \leq \theta < \pi$.
We then investigate the minimum value of
the excitation frequency $\Omega(\theta)$ given by Eq.~(\ref{Omega}).
The
collective oscillation will be dominated by the
compression mode for $0 < \theta < \pi/2$
and by the quadrupolar mode
for $\pi/2 < \theta < \pi$.
Moreover, $\theta=\pi/2$ represents a radial mode, and $\theta=0$
an axial mode.
The natural monopole and quadrupole operators
correspond to $\theta=\pi/4$ and
$\theta=\pi-\arctan(1/2) \approx 0.85\pi$,
respectively.

\begin{figure}
\begin{center}
\includegraphics[width=130mm]{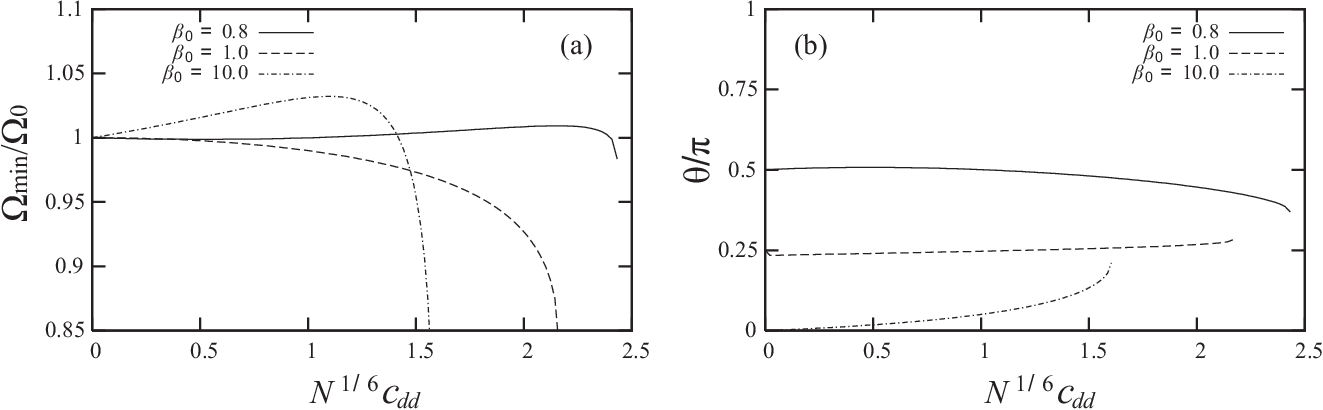}
\end{center}
\caption{
\label{fig-exmin-b0}
The minimum excitation frequency $\Omega_{\rm min}$ (a) and the angle $\theta$ that minimizes $\Omega(\theta)$ (b) as functions of the interaction strength
$N^{1/6}c_{dd}$ up to the critical values against instability.
$\Omega_{\rm min}$ is normalized to $\Omega_0$, the corresponding frequency
for the non-interacting system at each $\beta_0$:
$\Omega_0/\omega=1.7$, 2.0 and 0.2 for $\beta_0 =0.8$, 1.0 and 10, respectively.
The critical values are $N^{1/6}c_{dd}=2.433$, 2.166 and 1.603 for $\beta_0=0.8$, 1.0 and 10, respectively, see Figure~\ref{stability}.
}
\end{figure}
Figure \ref{fig-exmin-b0}(a)
shows the minimum excitation frequency $\Omega_{\rm min}$ as a function of the
interaction strength $N^{1/6}c_{dd}$
up to the critical value, while Figure \ref{fig-exmin-b0}(b) shows the angle $\theta$ that minimizes $\Omega(\theta)$.
For the spherical trap with $\beta_0=1.0$, the excitation frequency decreases monotonically
as the interaction strength increases, and the minimum-energy mode is the monopole mode.
For the prolate trap with $\beta_0=10.0$, the minimum-energy mode is dominated by the axial mode with $\theta \approx 0$ as the axial axis represents the direction of the soft confinement. Similarly, for the oblate trap with $\beta_0=0.8$, the minimum-energy mode is dominated by the radial mode with $\theta \approx \pi/2$ as the radial direction now becomes the soft axis. However, as the interaction strength increases towards the critical value, in both of these cases, the minimum-energy mode shifts towards the monopole mode, and we clearly see the tendency of the softening of the collective mode, indicating the approaching of the collapse instability. We note that, in particular for the case of
$\beta_0=0.8$,
$\Omega_{\rm min}$ does not completely decrease to zero at the critical value.
This could be due to the calculation of the average frequency of the collective oscillation by sum-rule method.
Deeper insights into collective excitations may be obtained from microscopic approaches such as the random-phase approximation~\cite{rs00,smsy02}.

%

\section{Expansion dynamics\label{sec-expansion}}
We now turn to the expansion dynamics of an initially trapped dipolar Fermi gas. This study is important as in most cold atom experiments, the atomic cloud is imaged after a period of free expansion. Furthermore, the expansion dynamics may bear the signature of the underlying interaction. The dipolar effects in chromium condensate are first observed in the expansion dynamics \cite{Cr,gps06,fesh}.

Our starting point is the Boltzman-Vlasov equation:
\begin{eqnarray}
\frac{\partial f\left( \mathbf{r},\mathbf{k},t\right) }{\partial t}+
\left(\frac{\hbar
\mathbf{k}}{m}+\frac{1}{\hbar}\frac{\partial}{\partial
\mathbf{k}}U\left( \mathbf{r},\mathbf{k}%
,t\right)\right)
\cdot \frac{\partial }{\partial \mathbf{r}}f\left( \mathbf{r},%
\mathbf{k},t\right) \nonumber \\
\;\;- \frac{1}{\hbar}\frac{\partial }{\partial \mathbf{r}}U\left( \mathbf{r},\mathbf{k}%
,t\right) \cdot \frac{\partial }{\partial \mathbf{k}}f\left( \mathbf{r},%
\mathbf{k},t\right) =0 \, ,  \label{bveo}
\end{eqnarray}%
where the effective potential $U$ includes both the external harmonic trap potential $U_{\rm ho}$ and the mean-field potential due to the dipole-dipole interaction:
\begin{eqnarray}
\fl
U\left( \mathbf{r},\mathbf{k},t \right) \! &=&\!U_{\mathrm{ho}}\left( \mathbf{%
r}\right) +\!\int \! d^3r^{\prime } \,n({\bf r}',t) \,V_{dd} \left( {\mathbf{r}}-{\mathbf{r}}%
^{\prime }\right) 
-\int \!\frac{d^3k^{\prime }}{(2\pi  )^{3}}\widetilde{%
V}_{dd}\left( {\mathbf{k}}-{\mathbf{k}}^{\prime }\right) f\left(
\mathbf{r},\mathbf{k}^{\prime },t\right)\,, \label{Ueff}
\end{eqnarray}%
where $\widetilde{
V}_{dd}( {\mathbf{k}})=(4\pi d^2/3)(3k_z^2/k^2-1)$ is the Fourier transform of $V_{dd}({\bf r})$. Note that the ${\bf k}$-dependence of the effective potential $U$ originates exclusively from the contribution of the exchange interaction, i.e., the last term at the r.h.s. of Eq.~(\ref{Ueff}). 

To study the dynamics, we shall make use of the scaling transformation
\begin{eqnarray*}
f({\mathbf{r}},{\mathbf{k}},t)=f_{0}({\mathbf{R}}(t),{\mathbf{K}}(t))\,, \\
R_{i}(t)=r_{i}/b_{i}(t) \,, \quad K_{i}(t)=b_{i}(t)k_{i}-m\dot{b}_{i}(t)r_{i}/\hbar \,, \quad (i=x,y,z)
\end{eqnarray*}
where $f_{0}$ represents the equilibrium Wigner distribution function obtained in previous section, whose form is given by Eq.~(\ref{psdtf}), and $b_{i}$ are the dimensionless scaling parameters. This scaling approach has been used previously to study the expansion of Fermi gases \cite{expF1,expF2} and Bose-Fermi mixtures \cite{expBF}.

From the Boltzman-Vlasov equation, we can derive the equations
governing the scaling parameters~\cite{expF1,expBF}
\begin{eqnarray}
\ddot{b}_{j}+\gamma_{j}^{2}b_{j}-\frac{\gamma _{j}^{2}}{b_{j}^{3}}+\frac{%
\epsilon _{dd}}{\langle R_{j}^{2}\rangle }\left[ \frac{\mathcal{T}_{j}({%
\mathbf{b},\dot\mathbf{b}})}{b_{j}}-\frac{\mathcal{T}_{j}({\mathbf{1}},\mathbf{0})}{b_{j}^{3}}\right]
=0, \label{dynb}
\end{eqnarray}
with $\gamma_j=\omega_j/\omega$, $\epsilon_{dd}=N^{1/6}c_{dd}$ and $\langle R_{j}^{2}\rangle =\int d^3R \,R_{j}^{2}n_{0}({\mathbf{R}})$ with $n_{0}$ being the equilibrium density. The second and third terms in Eq. (\ref{dynb}) represent, respectively, the restoring force and the kinetic energy. Collecting all contributions from interaction we have
\begin{eqnarray}
\mathcal{T}_{j}({\mathbf{b}},\dot{\mathbf{b}})&=&\int d^3R d^3R^{\prime }\,R_{j}%
\mathcal{W}({\mathbf{b}};{\mathbf{R}}-{\mathbf{R}}^{\prime })n_{0}({\mathbf{R%
}})\frac{\partial n_{0}({\mathbf{R}}^{\prime })}{\partial R_{j}^{\prime }}\nonumber\\
&&+\int d^3R d^3K d^3K^{\prime } \,R_{j}K_{j}\, \widetilde{\mathcal{%
W}}\left( {\mathbf{b}};{\mathbf{K}}-{\mathbf{K}}^{\prime }\right)   \nonumber \\
&& \times \sum_{i}\frac{\partial f_{0}({\mathbf{R}},{\mathbf{K}})}{\partial
K_{i}}\left[ \frac{\partial f_{0}({\mathbf{R}},{\mathbf{K}}^{\prime })}{%
\partial R_{i}}-\frac{b_{i}\dot{b}_{i}}{2\pi}\frac{\partial f_{0}({\mathbf{R}},{\mathbf{K}%
}^{\prime })}{\partial K_{i}^{\prime }}\right]\nonumber\\
&&+\int
d^3Rd^3Kd^{3}K^{\prime
}R_{j}K_{j}f_{0}\left(
\mathbf{R},\mathbf{K}^{\prime }\right)\nonumber \\
&&\times\sum_{i}\frac{\partial \widetilde{\mathcal{W}}\left(
{\mathbf{b}};{\mathbf{K}}-{\mathbf{K}}^{\prime }\right)}{\partial K_{i}}    \left[-\frac{\partial f_{0}({\mathbf{R}},{\mathbf{K}})}{%
\partial R_{i}}+\frac{b_{i}\dot{b}_{i}}{2\pi }\frac{\partial f_{0}({\mathbf{R}},{\mathbf{K}})}{\partial
K_{i}}\right],\label{cale}
\end{eqnarray}
where $\mathcal{W}({\mathbf{b}};{\mathbf{R}})=\frac{%
b_{x}^{2}X^{2}+b_{y}^{2}Y^{2}-2b_{z}^{2}Z^{2}}{%
(b_{x}^{2}X^{2}+b_{y}^{2}Y^{2}+b_{z}^{2}Z^{2})^{5/2}}$ is the dipole-dipole
interaction potential under the scaling transformation and $\widetilde{\mathcal{W%
}}({\mathbf{b}};{\mathbf{K}})$ represents its Fourier transform. Given the Wigner function in Eq.~(\ref{psdtf}), we obviously have $b_x=b_y=b_\rho$ as the free expansion will not change the cylindrical symmetry. Moreover, the integrations for terms involving $\dot b_i$ in Eq.~(\ref{cale}) vanish, so that ${\cal T}_j$ reduces to a function of ${\bf b}$ only. The analytical expressions for $\mathcal{T}_{j}({\mathbf b})$ can be found as
\begin{eqnarray}
{\cal T}_\rho({\mathbf b})&=&qb_{\rho }^{-2}b_{z}^{-1}\left[d_{\rho }(\beta^{-3/2}b_{\rho }/b_{z} )-d_{\rho }(\alpha^{-3/2}b_{\rho }/b_{z} )\right],\nonumber\\
\mathcal{T}_{z}({\mathbf{b}})&=&qb_{\rho }^{-2}b_{z}^{-1}\left[d_{z}(\beta^{-3/2}b_{\rho }/b_{z} )-d_{z}(\alpha^{-3/2}b_{\rho }/b_{z} )\right],
\end{eqnarray}
where $q=1024(3\lambda ^{3})^{1/2}/(2835\pi ^{2})$ and the functions
$d_j$ are defined as
\begin{eqnarray*}
d_{\rho}(x) &=& (1-x^2)^{-2}\left[2-7x^2-4x^4+9x^4g(x)\right],\\
d_{z}(x) &=& 2(1-x^2)^{-2}\left[1+10x^2-2x^4-9x^2g(x)\right],
\end{eqnarray*}
with $g(x)\equiv \mathrm{tanh}^{-1}\sqrt{1-x^2}/\sqrt{1-x^2}$. We note that $d_j$ are all monotonically decreasing functions of $x$ and bounded between $2$ and $-4$ for $x\in [0,\infty)$.

Here we focus on the time evolution of the atomic cloud aspect ratios in real and momentum spaces which are defined, respectively, as
\[ \kappa_r (t) =\sqrt{\frac{\langle r_{\rho }(t)^{2}\rangle }{\langle
r_{z}(t)^{2}\rangle }}\,,\;\mbox{ and }\;\kappa _{p}(t)=\sqrt{\frac{\langle p_{\rho }(t)^{2}\rangle }{\langle
p_{z}(t)^{2}\rangle }}\,, \]
where initially the system is prepared in the ground state inside the external trap. Straightforward calculations yield that
\begin{eqnarray*}
\kappa_r(t) &=& \beta^{-3/2}b_{\rho }(t)/b_{z}(t) \\
 \kappa _{p}(t) &=& \left[ \frac{\beta \lambda ^{2}\alpha ^{3}b_{\rho
}^{-2}+\alpha ^{2}\dot{b}_{\rho }^{2}}{\beta \lambda ^{2}b_{z}^{-2}+\alpha
^{2}\beta ^{3}\dot{b}_{z}^{2}}\right] ^{1/2}
\end{eqnarray*}
The initial cloud aspect ratios are determined by the ground state Wigner function and can be easily shown to be $\kappa_r(0) = \beta^{-3/2}$ and $\kappa_p (0) = \alpha^{3/2}$. To study the expansion dynamics, we turn off the trapping potential at $t=0$ and the cloud starts to expand. We then solve for $b_\rho(t)$ and $b_z(t)$ using Eq.~(\ref{dynb}) with the restoring force term $\gamma_j^2 b_j$ removed and with the initial conditions $b_\rho(0)=b_z(0)=1$.  Before presenting our results, we recall that when the exchange interaction is ignored, the direct dipolar interaction always tends to stretch the cloud along the direction of dipole moments in both real and momentum spaces~\cite{expF2}.

\begin{figure}
\begin{center}
\includegraphics[width=155mm]{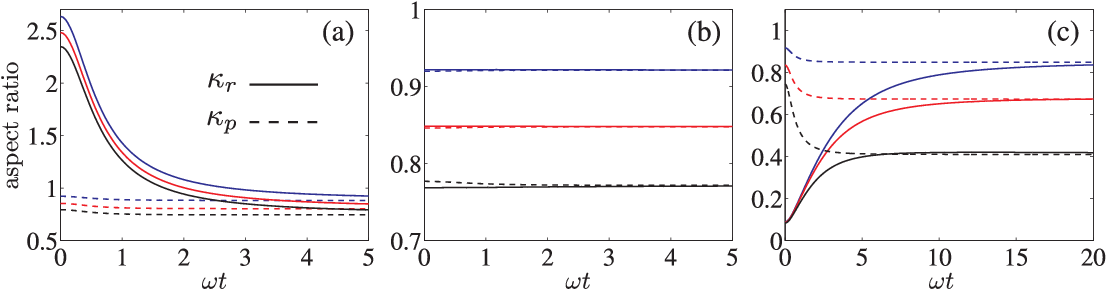}
\end{center}
\caption{
\label{ARt}
Cloud aspect ratio during time of flight in both momentum space (dashed lines) and real space (solid lines) for $\beta_0=0.5$ (a), $1$ (b), and $5$ (c). In each figure, the dipolar interaction strength are $N^{1/6}c_{dd} =0.5$, $1.0$, and $1.5$, in descending order.}
\end{figure}

Figure \ref{ARt} displays several examples of the cloud aspect ratio during time of flight for different trap geometries. As expected, asymptotically, the aspect ratios in momentum and real spaces become equal to each other, i.e., $\kappa_r(\infty) = \kappa_p(\infty) = \kappa_\infty$. A notable feature is that, regardless of the initial trap geometry, the shape of the expanding cloud eventually becomes prolate as $\kappa_\infty <1$. This result is in stark contrast to the expansion dynamics of a dipolar condensate whose asymptotic aspect ratio is sensitive to the initial trap geometry \cite{Cr,gps06,fesh}. Furthermore, the interaction effects during the time of flight is also evident from Figure~\ref{ARt}: Had interaction been ignored, the expansion would have become ballistic with $\kappa_p$ a constant in time. Figure~\ref{ARt}(b) indicates that the expansion is essentially ballistic for an initial spherical trapping potential, as for such traps, the interaction energy is rather weak as shown in Fig.~\ref{stability}(b).

\begin{figure}
\begin{center}
\includegraphics[width=120mm]{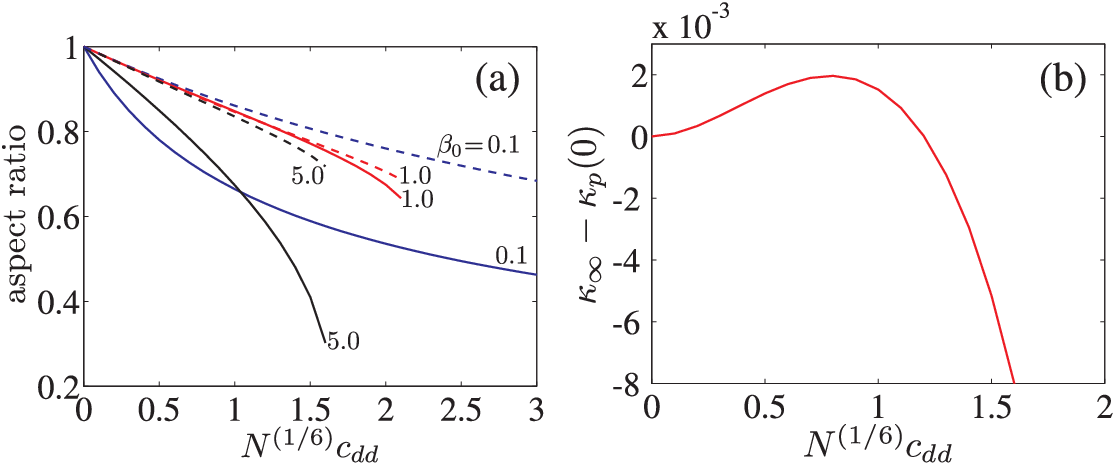}
\end{center}
\caption{
\label{AR}
(a) The dipolar interaction strength dependences of asymptotic aspect ratios $\kappa_\infty$ (solid lines) and the initial momentum space aspect ratio $\kappa_p(0)$ (dashed lines) for various trap aspect ratio $\beta_0$'s. (b) The difference between the asymptotic aspect ratio and the initial momentum space aspect ratio for $\beta_0=1$.
}
\end{figure}

That the expanded cloud eventually becomes prolate in shape is also obtained in Ref.~\cite{expF2} when the exchange dipolar interaction is ignored, which indicates that the effect of the exchange interaction during the expansion is not very important. This is consistent with Fig.~\ref{stability}(b) which shows that, except for nearly spherical traps, the magnitude of the direct energy is in general much larger than that of the exchange energy. However, we want to emphasize that the exchange term is crucial for the equailibrium momentum distribution inside the trap: Without the exchange term, the momentum distribution would be isotropic for any trap geometry.
To get a closer look, we compare in Fig.~\ref{AR} $\kappa_\infty$ with the initial momentum space aspect ratio $\kappa_p(0)$ which characterizes the momentum distribution for the ground state in the trap. The initial momentum distribution is always prolate in shape as $\kappa_p(0)<1$. In general, the effect of the interaction during the expansion, with the dominant contribution from the direct term, is to further enhance this anisotropy such that $\kappa_\infty<\kappa_p(0)$. Exceptions may occur for nearly spherical traps, for which one may have $\kappa_\infty>\kappa_p(0)$ as shown in Fig.~\ref{AR}(b). However, this effect is very small since, as we have already mentioned earlier, the total dipolar interaction is weak for such traps.

\section{Summary\label{sec-summary}}

In summary, we have studied the properties of dipolar Fermi gases
in both homogeneous system and in the cylindrical harmonic trap with
the dipole moments oriented along the symmetry axis.
The total energy functional of this system is
derived under the Hartree-Fock approximation.
The one-body density matrix in the energy functional is obtained from a variational ansatz based
on the Thomas-Fermi density distribution in phase-space representation, which accounts for the interaction-induced deformation in both real and momentum space. Our calculations show that deformation of the spatial density distribution
comes from the Hartree direct energy term, while
deformation of the momentum density distribution
arises from the Fock exchange energy term.
Note that the exchange term, a consequence of the anti-symmetry of the many-body fermionic wave function, does not appear in Bose-Einstein condensate.

We have calculated several thermodynamic quantities such as the pressure, the compressibility and the chemical potential of the homogeneous system and   investigated the low-lying collective excitations of a trapped dipolar Fermi gas using the sum rule method for various trap geometry and interaction strengths.
We observe the softening of the collective excitations as the interaction strength approaches the critical value for collapse.

%
%
%
%

Finally, we have studied the expansion dynamics of the initially trapped system. We show that, in stark contrast to dipolar condensate \cite{Cr,gps06,fesh}, the atomic Fermi gas will eventually become elongated along the direction of the dipoles regardless of the initial trap geometry. This feature makes it convenient to detect the dipolar effects in Fermi gases.


\ack
T.S. is supported by the DFG grant No. RO905/29-1.
S. Y. is supported by NSFC (Grant No. 10674141), National 973 program of China (Grant. No. 2006CB921205), and the ``Bairen'' program of Chinese Academy of Sciences. H.P. acknowledges
support from NSF, the Welch Foundation (Grant No.
C-1669), and the W. M. Keck Foundation.

\end{document}